\documentclass[12pt,a4,portrait]{article}
\usepackage{latexsym}
\usepackage{amssymb}
\usepackage{hyperref}
\usepackage{graphicx}
\usepackage{relsize}
\usepackage{anyfontsize}
\usepackage{mathtools}
\usepackage{multirow}
\usepackage[nottoc]{tocbibind}

\usepackage{amsmath, amsthm, amssymb}

\newcommand{\be}{\begin{equation}}
\newcommand{\ee}{\end{equation}}
\newcommand{\ben}{\begin{equation*}}
\newcommand{\een}{\end{equation*}}
\newcommand{\beq}{\begin{eqnarray}}
\newcommand{\eeq}{\end{eqnarray}}
\newcommand{\beqn}{\begin{eqnarray*}}
\newcommand{\eeqn}{\end{eqnarray*}}

\begin{document}

\title{Parametrizations in scalar-tensor theories of gravity and the limit of general relativity}
\author{Laur J\"arv\thanks{laur.jarv@ut.ee}, Piret Kuusk\thanks{piret.kuusk@ut.ee},
Margus Saal\thanks{margus.saal@ut.ee}, Ott Vilson\thanks{ovilson@ut.ee} \\
Institute of Physics, University of Tartu, Ravila 14c, Tartu 50411, Estonia} 

\date{}
\maketitle

\begin{abstract}
We consider a general scalar-tensor theory of gravity and
review briefly different forms it can be presented
(different conformal frames and scalar field parametrizations).
We investigate the conditions under which its field equations
and the parametrized post-Newtonian parameters
coincide with those of general relativity.
We demonstrate that these so-called limits of general relativity
are independent of the parametrization of the scalar field,
although the transformation between scalar fields may be
singular at the corresponding value of the scalar field.
In particular, the limit of general relativity can
equivalently be determined and investigated
in the commonly used Jordan and Einstein frames.
\end{abstract}

\section{Introduction}\label{Introduction}

The still unknown nature of the phenomena of dark matter and dark energy facilitates continued interest in the alternatives to Einstein's general relativity (for a comprehensive review see e.g. Ref.~\cite{Clifton_review}).
A simple and straightforward extension of general relativity (GR) is provided by the Jordan-Brans-Dicke theory \cite{jordan, brans} and its generalization scalar-tensor gravity (STG) \cite{bergmann, wagoner},
where an additional scalar field participates in the gravitational interaction.
As was proposed by Dicke, it is possible via a conformal rescaling of the metric (frame change) and reparametrization of the scalar field to transform the STG action into another representation,  equivalent to the original one if the units of measurement are also appropriately rescaled \cite{dicke}.    
However, the precise significance of this transformation and the 
physical and mathematical equivalence of different representations are 
still a topic for an ongoing debate 
(for a glimpse of the most recent papers see e.g. Refs.~\cite{Quiros:2011wb, Tsamparlis:2013aza, Chiba:2013mha, Stabile:2013eha}).
The present work contributes to the discussion by  following our earlier study \cite{meie_einstein} in asking what happens 
when the transformation from one
parametrization of the scalar field to another
is singular.

In the following section we write down the most general STG action involving four free functions, the transformations that leave this action invariant, and the ensuing field equations. In Sec.~\ref{Some widely used action functionals} we recall some of the most used STG frames and parametrizations, and collect some useful relations between them. Then Sec.~\ref{Field equations and the limit of general relativity} argues that the limit that reduces the STG field equations into those of general relativity does not depend on the particular frame and parametrization, in particular, it is not affected by a possible singularity in reparametrizing the scalar field. 
Sec.~\ref{Parametrized post-Newtonian approximation} gives the parametrized post-Newtonian (PPN) parameters for STG in its most general form as well as for the most used special cases introduced before, in order to witness again that the conditions for these parameters to coincide with those of GR are independent of the parametrization.
Therefore if in some parametrization there is a certain value of the scalar field that takes STG to its GR limit, there is  necessarily a corresponding value (or values) of the reparametrized field, that does the same in another frame and parametrization. This conclusion is illustrated by an example in Sec.~\ref{Example} and summarized in Sec.~\ref{Conclusion}.

\section{General action functional and field equations}\label{General action functional and field equations}

The most general action functional for a scalar-tensor theory of gravity 
 including scalar self-interaction only through scalar field but not its derivatives
 was written down by Flanagan \cite{Flanagan},
\be
\label{fl_moju}
S = \frac{1}{2\kappa^2}\int_{V_4}d^4x\sqrt{-g}\left\lbrace {\mathcal A}(\Phi)R-
{\mathcal B}(\Phi)g^{\mu\nu}\nabla_\mu\Phi \nabla_\nu\Phi - 2\kappa^2V(\Phi)\right\rbrace 
+ S_m\left(e^{2\alpha(\Phi)}g_{\mu\nu},\chi\right) \,.
\ee
It contains four arbitrary functions of the dimensionless scalar field $\Phi$: nonminimal coupling
function ${\mathcal A}(\Phi)$, generic kinetic coupling of the scalar field ${\mathcal B}(\Phi)$,
self-interaction potential of the scalar field $V(\Phi)$ and conformal coupling $e^{2\alpha(\Phi)}$
between the metric $g_{\mu\nu}$ and matter fields $\chi$. Note that ${\mathcal A}(\Phi)$, ${\mathcal B}(\Phi)$ and
$\alpha(\Phi)$ are dimensionless, but for the convenience of notation in cosmology the scalar potential is assumed to be of the dimension
of energy density, $[V]=[\rho]$ in units $c = 1$. If we impose a physical condition that
gravitational interaction is always finite and attractive, the nonminimal coupling function must
satisfy  $0 < {\mathcal{A}} < \infty$. We also assume from physical considerations
 that self-interaction potential is non-negative,  $0 \leq V(\Phi) < \infty  $. 

As demonstrated by Flanagan \cite{Flanagan}, two of the four arbitrary functions can 
be fixed by transformations that 
contain two arbitrary functions $\bar{\gamma}(\bar{\Phi})$, $\bar{f}(\bar{\Phi})$ and
leave the structure of action functional (\ref{fl_moju}) invariant:  
\beq
\label{fl_teisendused}
	g_{\mu\nu} &=& e^{2\bar{\gamma}(\bar{\Phi})}\bar{g}_{\mu\nu} \,,	 \\
	\Phi &=& \bar{f}(\bar{\Phi}) \,.	
\eeq
We will call the first transformation the change of the frame and the second one the
reparametrization of the scalar field. The change of the frame is in fact a conformal
rescaling of the metric and we assume that it is reasonable, i.e. the function
$\bar{\gamma}(\bar{\Phi})$ and its derivative $d\bar{\gamma}/d\bar{\Phi}$ do not diverge at any $\bar{\Phi}$. 

The transformed action functional (\ref{fl_moju}) retains its form
\be \label{fl_teisendatud_moju}
 {\bar S} = \frac{1}{2\kappa^2}\int_{V_4}d^4x\sqrt{-{\bar g}}
 \{ {\bar {\mathcal A}}({\bar \Phi}){\bar R}-{\bar {\mathcal B}}({\bar \Phi})
{\bar g}^{\mu\nu}{\bar \nabla}_\mu{\bar \Phi}{\bar \nabla}_\nu{\bar \Phi} - 2\kappa^2{\bar V}
({\bar \Phi}) \} + S_m\left(e^{2{\bar \alpha}({\bar \Phi})}{\bar g}_{\mu\nu},\chi\right) \,,
\ee
with transformed functions \cite{Flanagan}
\be
\label{fl_fnide_teisendused}
\begin{array}{rcl}
	\bar{\mathcal{A}}(\bar{\Phi}) &=& e^{2\bar{\gamma}(\bar{\Phi})}
	{\mathcal A} \left( {\bar f}( {\bar \Phi})\right) \,,\\
	{\bar {\mathcal B}}({\bar \Phi}) &=& e^{2{\bar \gamma}({\bar \Phi})}\left( 
	{\mathcal B}\left(\bar{f}(\bar{\Phi})\right)\left(\bar{f}^\prime\right)^2 -
	 6\left(\bar{\gamma}^\prime\right)^2{\mathcal A}\left(\bar{f}(\bar{\Phi})\right) -
	  6\bar{\gamma}^\prime\bar{f}^\prime \mathcal{A}^\prime \right) \,, \\
	\bar{V}(\bar{\Phi}) &=& e^{4\bar{\gamma}(\bar{\Phi})} \, V\left(\bar{f}(\bar{\Phi})\right) \,, \\
	\bar{\alpha}(\bar{\Phi}) &=& \alpha\left(\bar{f}(\bar{\Phi})\right) + \bar{\gamma}(\bar{\Phi})\,,
\end{array}
\ee
where $\bar{f}^\prime  \equiv \frac{d\bar{f}(\bar{\Phi})}{d\bar{\Phi}}$, $\mathcal{A}^\prime \equiv \frac{d \mathcal{A}(\Phi)}{d \Phi}$ etc.
Note that at the conformal transformation (\ref{fl_teisendused}) the energy-momentum tensor $T_{\mu \nu} = - \frac{2}{\sqrt{-g}}\frac{\delta S_m}{\delta g^{\mu \nu}}$ transforms as $\bar{T}_{\mu \nu} = e^{2 \bar{\gamma}} T_{\mu \nu}$ and its trace as $\bar{T} = e^{4 \bar{\gamma}} T$.

From these transformation rules we can notice the following.

\begin{itemize}
 
\item The conditions on nonminimal coupling function $0 < {\mathcal{A}}  < \infty $ and 
self-interaction potential $0 \leq V(\Phi) < \infty$ are
preserved, i.e.  $0 < {\bar{\mathcal{A}}}  < \infty$ and ${0 \leq \bar V}({\bar\Phi}) < \infty$.

\item If in some frame $\alpha =0 $, then in any other frame $|{\bar\alpha} | < \infty$.

\item If we want to avoid ghosts, i.e. if there is a frame where the tensorial and scalar part of the gravitational interaction are separated with ${\mathcal A}=1$ and ${\mathcal B}>0$, then in any related frame and parametrization it follows that 
\be \label{no_ghosts}
2 \bar{\mathcal{A}} \bar{\mathcal{B}} + 3 (\bar{\mathcal{A}}^{\prime})^2 >0 \,.
\ee
We assume this relation to hold.
\end{itemize}  


The field equations can be derived from the general action functional (\ref{fl_moju}) 
by varying with respect to metric tensor $g^{\mu\nu}$ and scalar field $\Phi$, respectively:
\beq
 {\mathcal A}G_{\mu\nu} + 
\left[\frac{1}{2}{\mathcal B}+{\mathcal A}^{\prime\prime}\right]g_{\mu\nu}\nabla_\rho\Phi\nabla^\rho\Phi 
- \left[{\mathcal B} + {\mathcal A}^{\prime\prime}\right]\nabla_\mu\Phi\nabla_\nu\Phi  \qquad && \nonumber  \\
\label{gr_vorrand} +{\mathcal A}^\prime\left[g_{\mu\nu}\Box\Phi - \nabla_\mu\nabla_\nu\Phi\right] 
+ \kappa^2g_{\mu\nu}V &=& \kappa^2T_{\mu\nu}\,, \\
\label{sk_vorrand} 
\frac{1}{2}R {\mathcal A}^\prime + \frac{1}{2}{\mathcal B}^\prime g^{\mu\nu}\nabla_\mu\Phi\nabla_\nu\Phi 
+ {\mathcal B}\Box\Phi - \kappa^2V^\prime &=& -\kappa^2\alpha^\prime T  \,.
\eeq
Upon substituting  the scalar curvature $R$ from the first equation into the second one and multiplying by $2 \mathcal{A}$,
the equation for the scalar field reads
 \be
\label{sk_vorr}
\displaystyle
\left( 2 {\mathcal A} {\mathcal B} + 3 (\mathcal{A}^{\prime})^2 \right) \Box\Phi + \frac{\left( 2 {\mathcal A} {\mathcal B} + 3 (\mathcal{A}^{\prime})^2  \right)^{\prime}}{2}
 g^{\mu\nu}\nabla_\mu\Phi\nabla_\nu\Phi
- 2 \kappa^2 \left(\mathcal{A} V^\prime - 2{\mathcal A}^{\prime}V\right) = \kappa^2 \left({\mathcal A}^{\prime} - 2 \mathcal{A} \alpha^{\prime} \right)T \,.
\ee
A direct calculation demonstrates that upon the transformation (\ref{fl_fnide_teisendused})
 all terms in the equation of the metric tensor (\ref{gr_vorrand})
acquire a common factor $e^{2 \bar{\gamma}}$, which we
have assumed to be regular. However, the transformed equation for the scalar field 
(\ref{sk_vorr}) gets a common factor $e^{6 \bar{\gamma}} \bar{f}^\prime$. 
If the transformation is regular, i.e. $\bar{f}^\prime \not= 0$, $\bar{f}^\prime \not= \infty$, this equation should yield an equivalent account of the same physics in different parametrizations. 
What happens for the points where $\bar{f}^\prime$ fails to be finite needs extra attention.
 
Finally, from the field equations (\ref{gr_vorrand}), (\ref{sk_vorrand}) a continuity equation
follows:
\be
\label{EI_jaavus}
\nabla_\mu T^{\mu\nu} = \alpha^\prime T \nabla^\nu\Phi \,.
\ee
If $\alpha^\prime = 0$ the right-hand side vanishes and the usual 
conservation of energy law holds; let us call the $\alpha = 0$ case the Jordan frame. 
Another well-known frame is the Einstein frame with ${\mathcal A} =1$ and in general  
$\alpha^\prime \not= 0$.  

\section{Some widely used action functionals}\label{Some widely used action functionals}

Sometimes in the literature one may encounter  treatments which fix the frame (i.e. fix $\alpha (\Phi)$, e.g. $\alpha = 0$), but leave the parametrization of the scalar field
unfixed, thus keeping three arbitrary functions in the STG action functional. 
But most often one meets a few distinct forms of the STG action functional obtained from  the general action (\ref{fl_moju}) by 
fixing two of the four arbitrary functions. These are the following.

\textit{1. The Jordan frame action in the Brans-Dicke-Bergmann-Wagoner parametrization} (JF BDBW) \cite{brans, bergmann, wagoner} for the 
scalar field $\Psi$ fixes ${\mathcal A} = \Psi$, $\alpha = 0$, while keeping ${\mathcal B} = \omega(\Psi)/\Psi$, $V= V(\Psi)$:
\beq \label{jf4da}
S  = \frac{1}{2 \kappa^2} \int d^4 x \sqrt{-g}
        	        \left[ \Psi R - \frac{\omega (\Psi ) }{\Psi}
        		\nabla^{\rho}\Psi \nabla_{\rho}\Psi
                  - 2 \kappa^2 V(\Psi)  \right] 
                  + S_m\left(g_{\mu\nu},\chi\right)  \,.
\eeq
The original Brans-Dicke gravity (JF BD) \cite{brans} with a potential is a special case where $\omega=const.$,
\beq \label{jf_bd}
S  = \frac{1}{2 \kappa^2} \int d^4 x \sqrt{-g}
        	        \left[ \Psi R - \frac{\omega}{\Psi}
        		\nabla^{\rho}\Psi \nabla_{\rho}\Psi
                  - 2 \kappa^2 V(\Psi)  \right] 
                  + S_m\left(g_{\mu\nu},\chi\right)  \,.
\eeq

\textit{2. The Jordan frame action in the parametrization used by Boisseau, Esposito-Far\`{e}se,  Polarski and Starobinsky} (JF BEPS) \cite{boisseau, EspositoFarese} for the scalar field as $\phi$ is obtained by taking ${\mathcal B} = 1$, $\alpha = 0$, while having ${\mathcal A} = F(\phi)$,  $V  = V(\phi)$:
\beq \label{EFP}
S  = \frac{1}{2 \kappa^2} \int d^4 x \sqrt{-g}
        	        \left[ F(\phi) R - 
        		\nabla^{\rho}\phi \nabla_{\rho}\phi
                  - 2  \kappa^2 V(\phi)  \right] 
                  + S_m\left(g_{\mu\nu},\chi\right)   \,.
\eeq
In the so-called nonminimal coupling case (JF nm),
the function $F$ has a distinct form  $F(\phi) = 1 - \xi \phi^2 $, where $\xi$ is a 
dimensionless parameter: 
\beq
S  = \frac{1}{2 \kappa^2} \int d^4 x \sqrt{-g}
        	          \left[ \left(1 - \xi \phi^2 \right)R - 
        		\nabla^{\rho}\phi \nabla_{\rho}\phi
                  - 2 \kappa^2 V(\phi)  \right] 
                  + S_m\left(g_{\mu\nu},\chi\right)  \,.
\eeq

\textit{3. The Einstein frame action in canonical parametrization} (EF can) \cite{dicke, bergmann, wagoner} for the scalar field denoted as $\varphi$, fixes
${\mathcal A} = 1$, ${\mathcal B} = 2$,
 while keeping $\alpha = \alpha(\varphi)$ and $V=V(\varphi)$:
\beq \label{ef_action}
S= 
 {1 \over 2 \kappa^2} \int d^4 x \sqrt{-g} 
\left[ R - 2 g^{\mu\nu} \, \nabla_{\mu} \varphi \nabla_{\nu} \varphi
- 2 \kappa^2 \, V(\varphi) 
\right] + S_{m}\left(e^{2\alpha(\varphi)}g_{\mu\nu},\chi\right)  \,.
\eeq
The well known Einstein gravity with minimally coupled scalar field (EF min) can be viewed as a special case here with $\alpha=0$,
\beq \label{ef_min_action}
S= 
 {1 \over 2 \kappa^2} \int d^4 x \sqrt{-g} 
\left[ R - 2 g^{\mu\nu} \, \nabla_{\mu} \varphi \nabla_{\nu} \varphi
- 2 \kappa^2 \, V(\varphi) 
\right] + S_{m}\left(g_{\mu\nu},\chi\right)  \,.
\eeq
However, in the latter case the scalar field equation (\ref{sk_vorr}) does not contain matter energy-momentum $T$ as a source and strictly speaking the scalar field is not mediating the gravitational interaction any more. 

The transformations between these most common frames and parametrizations are presented in Table~\ref{table1}.
Note that the mutual derivatives of the scalar field in different parametrizations 
included in the Table~\ref{table1} are in fact
just $\bar{f}^\prime$ which should satisfy the conditions $\bar{f}^\prime \not= 0$, 
$\bar{f}^\prime \not= \infty$
for a transformation to be regular.


\begin{table}[t!]
\centering \small
\begin{flushleft}
\begin{tabular}{llll} 
  & JF BDBW ($\Psi$) &  JF BEPS ($\phi$) &   EF can ($\varphi$) \\                                
\hline \hline \\
JF BDBW ($\Psi$) &
 Identity  & 
 $F(\phi) = \Psi$ & 
 $\alpha ( \varphi) = -\frac{1}{2}\ln \Psi$ 
\vspace{2mm} \\ 
   &    
   & 
  $(\frac{d\phi}{d \Psi})^2 = \frac{\omega (\Psi)}{\Psi}  $ & 
  $(\frac{d \varphi}{d \Psi})^2 =\frac{2 \omega(\Psi) + 3}{4 \Psi^2} $  \vspace{2mm}\\
   & 
   & 
  $(\frac{dF}{d\phi})^2 = \frac{\Psi}{\omega (\Psi)}$ &  
  $(\frac{d\alpha}{d\varphi} )^2 = \frac{1}{2 \omega(\Psi) + 3}$  
\vspace{6mm}\\
JF BD ($\Psi$) &
 $\omega=\mathrm{const.}$  & 
 $F(\phi) = \Psi$ & 
 $\alpha ( \varphi) = -\frac{1}{2}\ln \Psi$ 
\vspace{2mm} \\ 
 &    
   & 
  $(\frac{d\phi}{d \Psi})^2 = \frac{\omega}{\Psi}  $ & 
  $(\frac{d \varphi}{d \Psi})^2 =\frac{2 \omega + 3}{4 \Psi^2} $  \vspace{2mm}\\
   & 
   & 
  $(\frac{dF}{d\phi})^2 = \frac{\Psi}{\omega}$ &  
  $(\frac{d\alpha}{d\varphi} )^2 = \frac{1}{2 \omega + 3}$  
\vspace{4mm}\\
\hline \\
JF BEPS ($\phi$)& 
  $\Psi = F(\phi)$ & 
  Identity & 
  $\alpha (\varphi)  = -\frac{1}{2} \ln F (\phi)$ 
\vspace{2mm}\\
  &  
 $\frac{d\Psi}{d\phi} = \frac{dF}{d\phi} $ & 
  &  
 $(\frac{d\varphi}{d\phi})^2 = \frac{3}{4} (\frac{d \ln F(\phi)}{d\phi})^2 $        \vspace{2mm}\\    
  & 
 $\omega(\Psi) = F(\phi)\frac{1}{(\frac{dF}{ d\phi})^2}$ & 
  &
 $\qquad \quad + \frac{1}{2 F(\phi)}  $ 
\vspace{6mm}\\                                                                                                              
JF nm ($\phi$) & 
 $\Psi = 1 - \xi \phi^2 $ 
  & 
 $F (\phi) = 1 - \xi \phi^2$ & 
 $\alpha (\varphi)= \frac{1}{2} \ln \left(\frac{1}{ 1 - \xi \phi^2}\right) $ \vspace{2mm}\\
  & 
 $\frac{d\Psi}{d\phi} =  - 2\xi \phi $ & 
 & 
 $(\frac{d \varphi}{d \phi})^2 = \frac{1 - \xi \phi^2 + 6 \xi^2 \phi^2}{2(1 -  \xi \phi^2)^2} $   
\vspace{2mm}\\
  & 
 $\omega(\Psi) =  \frac{\Psi}{4 \xi (1 - \Psi)}=  \frac{1 - \xi \phi^2}{4 \xi^2 \phi^2} $ &
  &
 $\phi^2 = \frac{1}{\xi}(1 - e^{-2 \alpha (\varphi)})$      
\vspace{4mm}\\
\hline \\                                                 
EF can ($\varphi$) & 
 $\Psi = e^{-2 \alpha (\varphi)}$ &         
 $F(\phi) = e^{-2\alpha(\varphi)}$ & 
 Identity 
\vspace{2mm}\\
  &
 $(\frac{d\Psi}{d\varphi})^2 = 4 e^{-4 \alpha (\varphi)} (\frac{d\alpha}{d\varphi} )^2 $ &
$\left(\frac{d\phi}{d\varphi}\right)^2 = 2e^{-2\alpha(\varphi)} \times$ 
\vspace{2mm}\\
  &  
  &  \quad \quad $\times\left(1-3\left(\frac{d\alpha}{d\varphi}\right)^2\right) $
  & 
\vspace{2mm}\\
  &  
$\omega (\Psi) = \frac{1}{2} \left( \frac{1}{(\frac{d\alpha}{d\varphi})^2} -3\right)$ & 
$(\frac{d F}{d \phi})^2 = \frac{2e^{-2\alpha(\varphi)}
 \left(\frac{d\alpha}{d\varphi}\right)^2}{1-3\left(\frac{d\alpha}{d\varphi}\right)^2}$
   &
 \vspace{6mm} \\
EF min ($\varphi$) & 
 $\Psi = 1$ &         
 $F(\phi) =1$ & 
 $\alpha=0$
 \vspace{2mm}\\
 &  
  & &

   \vspace{4mm} \\
\hline
\hline 
\end{tabular}
\end{flushleft}
\label{types}
\caption{ Transformations between  frames and parametrizations\label{table1}}
\end{table}


\section{Field equations and the limit of general relativity}\label{Field equations and the limit of general relativity}

Let us investigate the conditions under which a  STG  coincides with GR. Since the latter
one does not involve a dynamical scalar field, a natural assumption is $\Phi = \mathrm{const.}$, $\nabla_{\mu} \Phi = 0$. However, this condition should be 
made consistent by requiring that the source term for the scalar field also vanishes, otherwise constant $\Phi$ can not be maintained. Rewriting the scalar field equation (\ref{sk_vorr}) as
\be
\label{sk_vorr_1}
\displaystyle
\Box\Phi + \frac{1}{2} \frac{\left( 2 {\mathcal A} {\mathcal B} + 3 (\mathcal{A}^{\prime})^2  \right)^{\prime}}{ 2 {\mathcal A} {\mathcal B} + 3 (\mathcal{A}^{\prime})^2 }
 g^{\mu\nu}\nabla_\mu\Phi\nabla_\nu\Phi
 = 
\kappa^2 \frac{ \left({\mathcal A}^{\prime} - 2 \mathcal{A} \alpha^{\prime} \right)T + 2 \left(\mathcal{A} V^\prime - 2{\mathcal A}^{\prime}V\right) }{  2 {\mathcal A} {\mathcal B} + 3 (\mathcal{A}^{\prime})^2 }\,
\ee
it becomes clear that the STG equations can concur with those of GR at the values of $\Phi$ where the term on the RHS of (\ref{sk_vorr_1}) vanishes. 
Given that $\mathcal{A}$ is everywhere regular there are several possibilities.

The first and most obvious case is realized for $\Phi_{\bullet}$ which
should simultaneously  satisfy
\be \label{GR_limit_1a}
\left( {\mathcal A}^\prime - 2 \alpha^\prime{\mathcal A} \right)|_{\Phi_{\bullet}} \, T = 0 
\ee
and
\be \label{GR_limit_1b}
\left(\mathcal{A} V^\prime - 2{\mathcal A}^{\prime}V\right)|_{\Phi_{\bullet}} =0 \,,
\ee
while $\mathcal{B}|_{\Phi_{\bullet}}$ is finite and nonvanishing. In addition, for the full compliance with GR the factors in front of the kinetic terms in the Einstein equation (\ref{gr_vorrand}) and scalar field equation (\ref{sk_vorr_1}) should remain regular, hence
${\mathcal A}^{\prime \prime}|_{\Phi_{\bullet}}$ and ${\mathcal B}^{\prime}|_{\Phi_{\bullet}}$ should not diverge. If the latter is not the case, then the STG does not allow a solution which behaves exactly as GR, but it may still be possible to have solutions which dynamically approach GR as a limiting process, provided 
${\mathcal A}^{\prime \prime}\nabla_\mu \Phi \nabla_\nu \Phi \rightarrow  0$ or/and $\frac{{\mathcal B}^{\prime}}{{\mathcal B}}\nabla_\mu \Phi \nabla_\nu \Phi \rightarrow  0$ as $\Phi \rightarrow \Phi_{\bullet}$.
The result is the GR Einstein equation with $V|_{\Phi_{\bullet}}$ effectively playing the role of the cosmological constant.
It is instructive to observe that for JF BEPS parametrization the condition (\ref{GR_limit_1a}), (\ref{GR_limit_1b}) translates into $F^{\prime}|_{\phi_{\bullet}}=0$, $V^{\prime}|_{\phi_{\bullet}}=0$, for the nonminimal coupling case into $\phi_{\bullet}=0$, $V^{\prime}|_{\phi_{\bullet}}=0$, and for the EF canonical parametrization into 
$\alpha^{\prime}|_{\varphi_{\bullet}} = 0$, $V^{\prime}|_{\varphi_{\bullet}}=0$. 
But in the JF BDBW parametrization the condition (\ref{GR_limit_1a}) can not be realized for general matter ($T \neq 0$) at all since ${\mathcal A}^{\prime} \equiv 1$.

The second possibility to make the RHS of (\ref{sk_vorr_1}) to vanish is by having a value $\Phi_{\star}$ for which 
\be \label{GR_limit_2}
\frac{1}{\mathcal{B}} \big|_{\Phi_{\star}} = 0, 
\ee
while ${\mathcal A}^{\prime}|_{\Phi_{\star}}$, $\alpha^\prime|_{\Phi_{\star}}$, and ${V}^\prime|_{\Phi_{\star}}$ do not diverge. 
An important difference with the previous case is that here we do not have a 
constant solution for the scalar field, but only a process of approaching to that value. For this process to correctly yield the GR, we must demand that ${\mathcal B} \nabla_\mu \Phi \nabla_\nu \Phi \rightarrow 0$ as $\Phi \rightarrow \Phi_{\star}$. 
In addition, if ${\mathcal B}^{\prime}|_{\Phi_{\star}}$ or ${\mathcal A}^{\prime \prime}|_{\Phi_{\star}}$ happen to be singular as well, only the solutions with 
$\frac{{\mathcal B}^{\prime}}{\mathcal B} \nabla_\mu \Phi \nabla_\nu \Phi \rightarrow 0$,
${\mathcal A}^{\prime \prime} \nabla_\mu \Phi \nabla_\nu \Phi \rightarrow 0$,  lead to GR as a limit. 
For a later remark we note that if the Einstein equation (\ref{gr_vorrand}) and the scalar field equation (\ref{sk_vorr_1}) converge to the GR limit at the same ``rate'', i.e. if ${\mathcal B} \nabla_\mu \Phi \nabla_\nu \Phi \propto \frac{{\mathcal B}^{\prime}}{\mathcal B} \nabla_\mu \Phi \nabla_\nu \Phi $, then $\frac{{\mathcal B}^{\prime}}{\mathcal B^2}$ is finite.
Among the particular forms of STG the condition (\ref{GR_limit_2}) can be only realized in JF BDBW parametrization where it translates into $\frac{1}{\mathcal{\omega}}|_{\Psi_{\star}} = 0$. In the JF BEPS and nonminimal, and EF canonical parametrizations the function $\mathcal{B}$ is fixed to a constant value which precludes (\ref{GR_limit_2}).

At first there seems to be also a third option to make the RHS of (\ref{sk_vorr_1}) to vanish by letting $\frac{1}{\mathcal{A}^{\prime}} = 0$. 
However by looking at the scalar field equation (\ref{sk_vorrand}) this case 
turns out to be problematic. Namely, for GR it is well known that spacetime curvature and matter energy momentum are proportional to each other, $R \propto T$. But if ${\mathcal{A}}^\prime \rightarrow \infty$ then finite $T$ would correspond to vanishing $R$, unless $\alpha^{\prime}$ also blows up. The latter would complicate the continuity equation (\ref{EI_jaavus}).
Hereby we restrict our attention to the cases where
${\mathcal{A}}^\prime$ and $\alpha^\prime$ are not singular at the same value of scalar field $\Phi$, 
and therefore to achieve a GR-like behaviour we do not consider this possibility. 
Furthermore, for the sake of mathematical simplicity we also leave aside the rather fine-tuned theories where the conditions (\ref{GR_limit_1a}), (\ref{GR_limit_1b}) and (\ref{GR_limit_2}) are realized together, or where both the numerator and denominator of the RHS term of (\ref{sk_vorr_1}) vanish simultaneously thus requiring a much more thorough analysis.

Now, given the rather different conditions (\ref{GR_limit_1a}), (\ref{GR_limit_1b}) and (\ref{GR_limit_2}), one is entitled to ask whether there is any connection between them. It is interesting to note that there is. Under the transformations (\ref{fl_fnide_teisendused}) the quantities
\be \label{some_invariants}
\frac{ \left( {\mathcal A}^{\prime} - 2 {\mathcal{A}} {\alpha}^{\prime} \right)^2 }{2 {\mathcal A} {\mathcal B} +  3 ({\mathcal{A}}^{\prime})^2 } \, 
\qquad {\mathrm{and}} \qquad
\frac{ \left( \mathcal{A} V^\prime - 2{\mathcal A}^{\prime}V \right)^2 }{ \mathcal{A}^4 \left(  2 {\mathcal A} {\mathcal B} + 3 (\mathcal{A}^{\prime})^2 \right) } \,
\ee
retain their form, i.e. do not acquire extra terms or common factors.
Therefore, under a generic transformation the condition that the RHS of (\ref{sk_vorr_1}) vanishes remains invariant, i.e. if some $\Phi$ in a certain frame and parametrization satisfies it, then the corresponding $\bar{\Phi}$
in another frame and parametrization will also satisfy it. Although, it is completely feasible that $\Phi_{\bullet}$ obeying (\ref{GR_limit_1a}), (\ref{GR_limit_1b}) may
get translated into $\bar{\Phi}_{\star}$ obeying (\ref{GR_limit_2}). 

\section{Parametrized post-Newtonian approximation}\label{Parametrized post-Newtonian approximation}

For being viable, the scalar-tensor theory of gravity must pass the tests on local scales,
e.g., give a good account of the motions in our solar system. A natural framework
for such a check is provided by the parametrized post-Newtonian (PPN) formalism
adapted to slow motions in a weak field.
To compare GR and STG there are two nonvanishing
PPN parameters  $\gamma$ and $\beta$. 
They both have value $1$ for Einstein's general relativity which is also favored by current observations.
For an STG, the PPN parameters can deviate from unity as they depend on the spatially asymptotic background value of the scalar field \cite{Nordtvedt1970, Damour_Esposito-Farese}.
When STG has a potential, the PPN parameters cease to be constants as they also acquire an extra dependence on the distance $r$ from the source \cite{Olmo_prl, Olmo_prd,Perivolaropoulos, meie_ppn}. It is useful to express the result in units
where the Newtonian potential $U_{N} = \frac{\kappa^2 M}{8 \pi r}$ is dimensionless,
while the dimensionless constant $G_{\mathrm{eff}}(\Phi,r)$
 modifies multiplicatively Newton's gravitational constant $G_N = \frac{\kappa^2}{8\pi}$ and
determines the Cavendish force. 

It is possible to translate the general results \cite{meie_ppn} from JF BDBW parametrization into a generic representation of ${\mathcal{A}}$, ${\mathcal{B}}$, ${V}$, ${\alpha}$ by using the transformations  (\ref{fl_teisendused}) and (\ref{fl_fnide_teisendused})
where $\bar{\mathcal{A}}=\Psi$,  $\bar{\mathcal{B}}=\frac{\omega(\Psi)}{\Psi}$, and $\bar{\alpha}=0$. It follows that $\bar{\gamma} = -{\alpha}$, $\bar{\gamma}^\prime = -{\alpha^\prime} \bar{f}^\prime$, while the other necessary quantity $\bar{f}^\prime$ can be expressed by taking the derivative of the first line of Eq.~(\ref{fl_fnide_teisendused}). 
Since the PPN ansatz assumes flat Minkowski spacetime in spatial infinity, the internal consistency requires $V=0$, $V^\prime=0$ (all values of the functions taken at the asymptotic value of the scalar field).
In terms of the constant related to the scalar field effective mass,
\be
m_{\Phi} = \kappa \sqrt{ \frac{2 \mathcal{A}}{e^{2 \alpha} \left(2 \mathcal{A}\mathcal{B} + 3 (\mathcal{A}^\prime)^2 \right) } \frac{d^2 V}{d \Phi^2} } \,,
\ee
the results are
\beq
G_{\mathrm{eff}} &=& \frac{e^{2 \alpha}}{\mathcal{A}} \left(1+ \frac{\left( \mathcal{A^\prime} - 2 \mathcal{A} \alpha^\prime  \right)^2 e^{-m_{\Phi}r}}{2 \mathcal{A}\mathcal{B} + 3 (\mathcal{A}^\prime)^2 } \right) \,, \\
\gamma-1 &=& -\frac{2 e^{2 \alpha} \left( \mathcal{A^\prime} - 2 \mathcal{A} \alpha^\prime \right)^2 e^{-m_{\Phi}r} }{G_{\mathrm{eff}} \mathcal{A} \left( 2 \mathcal{A}\mathcal{B} + 3 (\mathcal{A}^\prime)^2 \right)} \,, \\
\beta-1 &=& \frac{ \left( \mathcal{A^\prime} - 2 \mathcal{A} \alpha^\prime  \right)^2 \left[ \left(2 \mathcal{A}\mathcal{B} + 3 (\mathcal{A}^\prime)^2 \right)^\prime  (\mathcal{A^\prime} - 2 \mathcal{A} \alpha^\prime ) - 2(2 \mathcal{A}\mathcal{B} + 3 (\mathcal{A}^\prime)^2 ) \left( \mathcal{A^\prime} - 2 \mathcal{A} \alpha^\prime \right)^\prime \right] e^{-2 m_{\Phi}r} }{2 \, e^{-4 \alpha} \, G_{\mathrm{eff}}^2 \, \mathcal{A} \left( 2 \mathcal{A}\mathcal{B} + 3 (\mathcal{A}^\prime)^2 \right)^3} \nonumber \\
&& \qquad - \frac{ \left( \mathcal{A^\prime} - 2 \mathcal{A} \alpha^\prime \right)^2 m_{\Phi} r}{ e^{-4 \alpha} \, G_{\mathrm{eff}}^2 \left( 2 \mathcal{A}\mathcal{B} + 3 (\mathcal{A}^\prime)^2 \right)}\left[ e^{-m_{\Phi}r} \ln m_{\Phi}r + \ldots \right] \,.
\eeq
Among the number of $r$-dependent terms in the square brackets on the last line only the contribution that is leading for large $m_{\Phi}r$ is given. 
For different
frames and parametrizations the corresponding expressions can be found in Table~\ref{table2}. These can be deduced by specifying the functions in the formulas above,
or using the transformations (\ref{fl_fnide_teisendused}) and the information in Table~\ref{table1}.

\begin{table}[t!]
\centering 
\begin{flushleft}
\begin{tabular}{ll}

\hline \hline \\
JF BDBW & 
$G_{\mathrm{eff}}=\frac{1}{\Psi} \left( 1+ \frac{e^{-m_{\Psi}r}}{2\omega + 3} \right)$ \,, \qquad \qquad
$m_{\Psi}= \kappa\sqrt{\frac{2 \Psi}{2\omega(\Psi) + 3}\frac{d^2V}{d\Psi^2} }$  
\vspace{2mm} \\
&
$\gamma-1= -\frac{2e^{-m_\Psi r}}{G_{\mathrm{eff}} \Psi (2\omega + 3)}$ 
\vspace{2mm} \\
&
$\beta-1= \frac{ \frac{d\omega}{d\Psi}  e^{-2m_{\Psi}r} }{G_{\mathrm{eff}}^2 \Psi (2\omega + 3)^3} - \frac{ m_{\Psi} r}{ G_{\mathrm{eff}}^2 \Psi^2 (2\omega + 3)} \left[ e^{-m_{\Psi}r} \ln (m_{\Psi}r) + \ldots \right]  $ 
\vspace{6mm} \\

JF BEPS & 
$G_{\mathrm{eff}}=\frac{1}{F} \left( 1+ \frac{\left(\frac{d F}{d\phi}\right)^2 e^{-m_{\phi} r} }{2 F+3\left(\frac{d F}{d\phi}\right)^2} \right)$ \,,  \qquad
$m_{\phi}=\kappa\sqrt{\frac{2 F}{2 F+3\left(\frac{d F}{d\phi}\right)^2} \frac{d^2 V}{d\phi^2} }$
\vspace{2mm}\\
& 
$\gamma-1 = -\frac{2\left(\frac{dF}{d\phi}\right)^2 e^{-m_\phi r}}{G_{\mathrm{eff}} F \left( 2F + 3 \left(\frac{dF}{d\phi}\right)^2 \right)}$ 
\vspace{2mm}\\
&
$\beta-1=\frac{\left(\frac{dF}{d\phi}\right)^2 \left( \left(\frac{dF}{d\phi}\right)^2 - 2 F \frac{d^2F}{d\phi^2} \right) e^{-2 m_{\phi} r}}
{G_{\mathrm{eff}}^2 F \left( 2F+3\left(\frac{dF}{d\phi}\right)^2\right)^3}
- \frac{ \left(\frac{dF}{d\phi}\right)^2 m_\phi r}{G_{\mathrm{eff}}^2 F^2 \left(2F+3\left(\frac{dF}{d\phi}\right)^2 \right)} \left[e^{-m_\phi r} \ln(m_{\phi}r) + \ldots \right]
$

\vspace{6mm}\\

JF nm & 
$G_{\mathrm{eff}}=\frac{1}{1-\xi \phi^2} \left(1+ \frac{2 \xi^2 \phi^2 e^{-m_{\phi}r}}{1-\xi \phi^2 + 6\xi^2 \phi^2 } \right)$ \,, \qquad 
$m_{\phi}=\kappa\sqrt{\frac{2 (1-\xi \phi^2) }{1-\xi \phi^2 + 6\xi^2 \phi^2 } \frac{d^2 V}{d\phi^2} }$
\vspace{2mm}\\
&
 $\gamma-1 = -\frac{4 \xi^2 \phi^2 e^{-m_{\phi}r}}{G_{\mathrm{eff}} (1-\xi \phi^2) (1-\xi \phi^2 + 6\xi^2 \phi^2)}$ 
\vspace{2mm}\\ 
& 
 $\beta-1=\frac{2 \xi^3 \phi^2 e^{-2m_{\phi}r}}{(1-\xi \phi^2) (1-\xi \phi^2 + 6\xi^2 \phi^2)^3} - \frac{2 \xi^2 \phi^2 m_{\phi}r }{(1-\xi \phi^2) (1-\xi \phi^2 + 6\xi^2 \phi^2)} \left[e^{-m_\phi r} \ln(m_{\phi}r) + \ldots \right]$

\vspace{6mm}\\

EF can & 
$G_{\mathrm{eff}} = e^{2\alpha}\left(1+ \left(\frac{d\alpha}{d\varphi}\right)^2 e^{-m_\varphi r}\right)$ \,, \qquad \qquad
$ m_\varphi = \kappa\sqrt{\frac{1}{2 e^{2\alpha}}\frac{d^2V}{d\varphi^2}}$
\vspace{2mm}\\
&
$\gamma - 1 = -\frac{2 e^{2\alpha} \left(\frac{d\alpha}{d\varphi}\right)^2 e^{-m_\varphi r}}{G_{\mathrm{eff}}}$ 
\vspace{2mm}\\
&
$\beta-1= \frac{e^{4\alpha} \left( \frac{d \alpha}{d \varphi}\right)^2 \frac{d^2 \alpha}{d \varphi^2} e^{-2 m_\varphi r} }{2 G_{\mathrm{eff}}^2 }
- \frac{ e^{4\alpha}\left( \frac{d \alpha}{d \varphi}\right)^2 m_\varphi r}{G_{\mathrm{eff}}^2} \left[ e^{-m_{\varphi}r} \ln (m_{\varphi}r) + \ldots \right]  $

\vspace{6mm}\\

EF min & 
$G_{\mathrm{eff}} = 1$ \,, \qquad \qquad \qquad  \qquad \qquad \qquad
$ m_\varphi = \kappa\sqrt{\frac{1}{2}\frac{d^2V}{d\varphi^2}}$
\vspace{2mm}\\
&
$\gamma - 1 = 0$ 
\vspace{2mm}\\
&
$\beta-1= 0$

\vspace{4mm} \\
\hline
\hline 
\end{tabular}
\end{flushleft}
\caption{PPN parameters in different frames and parametrizations\label{table2}}
\end{table}

The conceptual difference with the previous section is that now we have a static configuration and the functions of the scalar field are taken at their spatially asymptotic values. 
However, the analysis of the limit where the STG PPN parameters coincide with those of general relativity, viz. $G_{\mathrm{eff}}=1$, $\gamma=1$, $\beta=1$ proceeds quite analogously. The first option is provided by the condition (\ref{GR_limit_1a}), where in addition $\mathcal{B}$ is finite and $\left( \mathcal{A^\prime} - 2 \mathcal{A} \alpha^\prime \right)^2 \left( \mathcal{A^\prime} - 2 \mathcal{A} \alpha^\prime \right)^\prime =0$. Note that
the twin condition (\ref{GR_limit_1b}) is automatically satisfied due to the PPN ansatz. The second option would be given by (\ref{GR_limit_2}),
with $\alpha^\prime$, $\alpha^{\prime\prime}$ not infinite, and $\frac{\mathcal{B}^\prime}{\mathcal{B}^3}=0$. Comparing the latter with the discussion in the previous section we may note that the condition on $\mathcal{B}^\prime$ to achieve the GR limit is marginally less strict in PPN than the one obtained from the equations of motion, i.e.~$\frac{\mathcal{B}^\prime}{\mathcal{B}^2}$ finite. (A similar observation in the case of cosmology was made in Ref.~\cite{meie5}). The third option is realized by giving the scalar field a very large effective mass, i.e.
\be \label{large_mass}
\frac{1}{m_{\Phi}} \Big|_{\Phi_{\blacklozenge}}  = \left( \frac{2 \kappa^2 \mathcal{A}}{e^{2 \alpha} \left(2 \mathcal{A}\mathcal{B} + 3 (\mathcal{A}^\prime)^2 \right) } \frac{d^2 V}{d \Phi^2} \right)^{-\frac{1}{2}} \Big|_{\Phi_{\blacklozenge}} = 0 \,.
\ee
However, in that case it is not so obvious what the corresponding condition arising from the general equations of motion would be.

\section{Example}\label{Example}

To have an illustration let us take a look at a specific simple example. Let the JF BDBW functions be given by
\be \label{example_BDBW}
\mathcal{A}=\Psi \,, \qquad \mathcal{B}=\frac{\omega(\Psi)}{\Psi}=\frac{3}{2(1-\Psi)} \,, \qquad V=\frac{1}{\left(\frac{1}{2}-\Psi \right)^2} \,, \qquad \alpha=0 \,.
\ee
The attactive gravitation condition ($\mathcal{A}>0$) and no ghosts condition (\ref{no_ghosts}) delimit $0 < \Psi \leq 1$. Recalling the discussions in Secs.~\ref{Field equations and the limit of general relativity} and \ref{Parametrized post-Newtonian approximation} we may find that the field equations
and PPN parameters reduce to those of general relativity in several different occasions. 
\begin{itemize}
\item The first is realized when (\ref{GR_limit_2}) holds, i.e. $\Psi_{\star} =1$, while $\frac{\mathcal{B}^\prime}{\mathcal{B}^2}=\frac{2}{3}$ is finite. Here the PPN parameters also reduce to their general relativity values, as expected, since $\frac{\mathcal{B}^\prime}{\mathcal{B}^3}=0$.
\item The second possibility to reduce the field equations to GR only occurs when the trace of matter energy-momentum tensor $T=0$ and the condition (\ref{GR_limit_1a}) does not apply. Then (\ref{GR_limit_1b}) is satisfied at
$\Psi_{\bullet} =\frac{1}{4}$. The PPN parameters, however, do not coincide with those of GR now.
\item Finally, it is possible to draw the PPN parameters into the GR values by satisfying (\ref{large_mass}) with an extremely massive scalar field, $\Psi_{\blacklozenge} =\frac{1}{2}$. But now the field equations do not agree with those of general relativity.
\end{itemize}

We can transform the theory from the BDBW parametrization with $\Psi$ into the BEPS parametrization with $\phi$ by using Table~\ref{table1}. Integrating
\be \label{transform_BDBW_BEPS}
\mp \frac{d \phi}{d\Psi} = \sqrt{ \frac{\omega(\Psi)}{\Psi} } = \sqrt{ \frac{3}{2(1-\Psi)} }
\ee
gives (neglecting the additive integration constant)
\be \label{example_BEPS_BDBW}
\pm \phi = \sqrt{ 6(1-\Psi) } \,, \qquad \Psi=1-\frac{1}{6}\phi^2 \,,
\ee
and we see it is actually the nonminimal coupling subclass of BEPS, where
\be \label{example_BEPS}
\mathcal{A}=F(\phi) = 1-\frac{1}{6}\phi^2 \,, \qquad \mathcal{B}=1 \,, \qquad V=\frac{1}{\left(\frac{1}{2}-\frac{\phi^2}{6} \right)^2} \,, \qquad \alpha=0 \,.
\ee
Note that $\Psi$ is mapped doubly to $\phi$, as $\Psi \in (0, 1]$ translates into $\phi \in (-\sqrt{6}, 0]$ and  $\phi \in [0, \sqrt{6})$. 
Again, there are several possibilities to achive the general relativity limit of the field equations and PPN parameters.
\begin{itemize}
\item First, the field equations reduce to the ones of GR when Eq.~(\ref{GR_limit_1a}), given by $\mathcal{A}^\prime=0$, and (\ref{GR_limit_1b}), given by $\mathcal{A} V^\prime - 2 \mathcal{A}^\prime V =0$, are satisfied. The only common solution is $\phi_{\bullet} = 0$. A glance to Table~\ref{table2} reveals that the PPN parameters also trivially fall into their GR limit. By a direct comparison via (\ref{example_BEPS_BDBW}) it becomes obvious that this value of $\phi$ corresponds to the first case in the BDBW case.
\item If matter $T=0$ and the condition (\ref{GR_limit_1a}) is not enforced, the condition (\ref{GR_limit_1b}) alone has also the solution $\pm \phi_{\bullet} = \frac{3}{\sqrt{2}}$. This does not lead the PPN parameters to their GR values. A direct check by (\ref{example_BEPS_BDBW}) tells that the corresponding case in the BDBW parametrization was the second one.
\item Last, when the scalar field acquires an extremely large mass by (\ref{large_mass}) at $\pm \phi_{\blacklozenge} = \sqrt{3}$ the PPN parameters reduce to those of GR, but the field equations do not. It corresponds to the third case above.
\end{itemize}

We may transform the same theory from JF BDBW parametrization into EF canonical parametrization by integrating
\be \label{transform_BDBW_EIN}
\mp \frac{d \varphi}{d \Psi} = \sqrt{ \frac{2 \omega(\Psi)+3}{4 \Psi^2} } = \sqrt{ \frac{3}{4 \Psi^2(1-\Psi)} } \,,
\ee
which gives (neglecting the additive integration constant)
\be \label{example_BDBW_to_EIN}
\pm \varphi = \sqrt{3} \, \mathrm{arctanh} \sqrt{ 1 - \Psi } \,, \qquad \Psi = 1 - \tanh^2 \frac{\varphi}{\sqrt{3}} \,.
\ee
The functions characterizing the frame and parametrization are
\be \label{example_EIN}
\mathcal{A}=1 \,, \quad \mathcal{B}=2 \,, \quad V=\frac{1}{\left(\frac{1}{2}-\tanh^2 \frac{\varphi}{\sqrt{3}} \right)^2 \left(1-\tanh^2 \frac{\varphi}{\sqrt{3}} \right)^2} \,, \quad \alpha=-\frac{1}{2} \ln \left( 1- \tanh^2 \frac{\varphi}{\sqrt{3}} \right) \,.
\ee
Alternatively, one may embark from the JF BEPS parametrization and integrate
\be \label{transform_BEPS_EIN}
\pm \frac{d \varphi}{d \phi} = \sqrt{\frac{3}{4} \left(\frac{d \ln F(\phi)}{d\phi}\right)^2 + \frac{1}{2 F(\phi)} } =
 \frac{1}{\sqrt{2} \left(1 - \frac{\phi^2}{6} \right)} 
\ee
to obtain (again, neglecting the additive integration constant)
\be \label{example_BEPS_to_EIN}
\pm \varphi = \sqrt{3} \, \mathrm{arctanh} \frac{\phi}{\sqrt{6}} \,, \qquad \pm \phi = \sqrt{6} \, \tanh \frac{\varphi}{\sqrt{3}} \,.
\ee
The mapping into EF canonical parametrization is again double for JF BDBW, as 
$\Psi \in (0, 1]$ translates into $\phi \in (-\infty,0]$ and $\phi \in [0, \infty)$, while JF BEPS $\phi \in (-\sqrt{6}, \sqrt{6})$ translates into $\phi \in (-\infty,\infty)$ and equivalently into $-\phi \in (-\infty,\infty)$ according to the sign in Eq.~(\ref{transform_BEPS_EIN}).
Analogously with the other parametrizations we can discuss the general relativity limit of the field equations and PPN parameters in three cases.
\begin{itemize}
\item When the conditions (\ref{GR_limit_1a}) and (\ref{GR_limit_1b}) both hold, i.e. $\alpha^\prime=0$ and $V^\prime = 0$, the value of the scalar field is $\varphi_{\bullet} = 0$. It takes the PPN parameters to their GR limit and by direct check using (\ref{example_BDBW_to_EIN}) and (\ref{example_BEPS_to_EIN}) one can conclude it corresponds to the first cases discussed above.
\item For absent or radiative matter with $T=0$ the condition (\ref{GR_limit_1a}) does not apply and (\ref{GR_limit_1b}) alone is also solved by $\pm \varphi_{\bullet}= \sqrt{3} \, \mathrm{arctanh} \left( \frac{\sqrt{3}}{2} \right)$. The PPN parameters differ from those of GR and it is straightforward to check that this value of $\varphi$ corresponds to the second cases above.
\item The scalar field mass diverges at $\pm \varphi_{\blacklozenge}=\sqrt{3} \, \mathrm{arctanh} \left( \frac{1}{\sqrt{2}} \right)$, satisfying (\ref{large_mass}) and reducing the PPN parameters to their GR values. The field equations still differ from those of GR and we recognize correspondence to the third cases described above.

\end{itemize}

We see that the derivative of the transformation function $\bar{f}^\prime$ relating different scalar field parametrizations gets singular for different values of the field. For a transformation from JF BDBW to BEPS (\ref{transform_BDBW_BEPS}) it is singular at $\Psi=1$, $\phi=0$, for a  transformation from JF BDBW to EF canonical (\ref{transform_BDBW_EIN}) it is singular at $\Psi=1$, $\varphi=0$ and $\Psi=0$, $\varphi=\pm \infty$, while for a  transformation from JF BEPS to EF canonical (\ref{transform_BEPS_EIN}) it is singular at $\phi=\pm\sqrt{6}$, $\varphi=\pm\infty$. 
The value $\Psi=0$, $\phi=\pm\sqrt{6}$, $\varphi=\pm\infty$ also makes the the conformal transformation  singular. But strictly speaking, this value is actually outside the range of the assumed validity of the theory, since for JF BDBW and BEPS it violates the attractive gravity assumption, while for EF canonical the infinite value of the field is arguably unphysical since $\alpha$ becomes singular.

So, it is only the singularity of transformation and the GR limit occuring at JF BDBW $\Psi=1$ that is possibly problematic. However, we saw that despite the transformation becoming singular the general relativity limit in terms of the field equations and PPN parameters, namely the first cases discussed above, does also occur in the corresponding JF BEPS value $\phi=0$ and EF  canonical parametrization value $\varphi=0$. 
It is interesting that in the JF BDBW parametrization this GR limit is realized by satisfying the condition (\ref{GR_limit_2}), while in the JF BEPS and EF canonical parametrizations it comes from the conditions (\ref{GR_limit_1a}), (\ref{GR_limit_1b}). 
This confirms the discussion in Secs.~\ref{Field equations and the limit of general relativity} and \ref{Parametrized post-Newtonian approximation} that the existence of the GR limit is invariant of the parametrization.

\section{Conclusion}\label{Conclusion}
We studied general scalar-tensor gravity involving four free functions in different conformal frames and scalar field parametrizations. We investigated its general relativity limits in the sense of field equations and the values of PPN parameters coinciding with those of general relativity. Despite the transformation of the scalar field from one representation to another may possess a singularity, it turned out that the existence of general relativity limits is independent of the parametrization.

\bigskip
{\bf Acknowledgements}

This work was supported by the Estonian Science Foundation
Grant No. 8837, by the Estonian Ministry for Education and Science
Institutional Research Support Project IUT02-27 and by the European Union through 
the European Regional Development Fund (Centre of Excellence TK114).

\bigskip

\end{document}